# Cavity optomechanical liquid level meter using a twin-microbottle resonator

Motoki Asano, Hiroshi Yamaguchi, and Hajime Okamoto

*Abstract*—Cavity optomechanical devices can be made to have good compatibility with optical fiber technology by utilizing fiber-based waveguides and cavities and can be used in high-performance optical sensor applications. Such optomechanical microsensors have a great potential for exploring the properties of liquids, such as density, viscosity, and masses of included nanoparticles. However, as yet, there is no cavity optomechanical architecture that can be used to sense the liquid's shape, e.g., liquid level. In this paper, we report a demonstration of a liquid-level meter using a twin-microbottle resonator that can make measurements at arbitrary positions and depths in the liquid. The twin-microbottle resonator has a maximum diameter of 68 μm and length of 800 μm. By immersing one part of it in water and keeping the other part in air, the mechanical radial breathing mode can be read out sensitively while maintaining a high optical quality factor of the optical whispering gallery mode regardless of the water immersion. This high mechanical displacement sensitivity provides a frequency resolution that is high enough to measure the mechanical frequency shift due to the water immersion and resolves the water level to 2.6±0.9 pm. This unique liquid-level meter based on a highly sensitive cavity optomechanical setup can be used to detect tiny fluctuations of various air-liquid and liquid-liquid interfaces.

*Index Terms*— Cavity optomechanics, whispering-gallery-mode microresonator,

## I. INTRODUCTION

CAVITY optomechanics is a promising way to achieve ultrasensitive measurements of mechanical motion by using a high-Q optical cavity [1]. It provides a high signal-noise ratio for displacement measurements, which in turn provide highly stable estimations of mechanical resonance frequencies. Moreover, such a stable frequency estimation capability enables us to perform cavity optomechanical sensing, i.e., measuring the frequency shift of mechanical resonance via the high-Q optical resonance. This technique has been utilized in various sensor applications, such as accelerometers [2], magnetometers [3],[4], chemical sensors [5], and biochemical sensors [6]-[9].

Recently, cavity optomechanical sensors have been studied for use as ultrasensitive chemical and biological sensors [9]-[14] and high-frequency rheometers [15] in liquid environments. To perform highly sensitive detection of mechanical motion in a liquid environment, it is of utmost importance to avoid light scattering and absorption loss in the liquid and to keep the optical Q factor high. The pioneering studies on cavity optomechanical sensing in liquids used a high-refractive-index microdisk with a liquid droplet [9]-[11] or silica microcapillary with fluid flowing inside the capillary [12]-[14]. Although these methods can be used to sense globally distributed targets inside a liquid, such as macroscopic viscosity or tiny dispersed particles, their fixed-by-design device architectures are not suitable for sensing spatially local information about the liquid. Here, we have recently reported a free-access liquid probe based on a twin microbottle cavity optomechanical resonator (TMBR) [16],[17]. This resonator enables not only high-sensitivity monitoring of local viscosity and density but also mass detection at arbitrary locations in a liquid while keeping a high optical Q factor.

In the study reported in this paper, we demonstrated another application of the TMBR: a highly sensitive liquid-level meter. The TMBR consists of two microbottle optomechanical resonators fabricated on a silica glass wherein high-Q optical whispering gallery modes (WGMs) are localized in each microbottle and high-Q mechanical radial breathing modes (RBMs) couple them. This fiber-type device structure can be used to make measurements at arbitrary positions in the liquid. By partially immersing one of the microbottles in the liquid, sensitive optical readout of the liquid level through the mechanical frequency shift due to the fluid-structure interaction becomes possible while maintaining a high-Q optical resonance.

## II. PRINCIPLE OF LIQUID LEVEL DETECTION

The setup for the liquid-level measurement using the TMBR is shown in Fig. 1(a). There are three interactions, the optomechanical interaction in the top microbottle, the mechanical interaction between the two microbottles, and the fluid-structure interaction in the bottom microbottle [Fig. 1(b)].

The optomechanical interaction between the optical WGM and mechanical RBM occurs in the top microbottle resonator. The total internal reflection in the optical WGM causes radiation pressure along the radial direction which efficiently couples to the mechanical RBM in the microbottle structure. This coupling allows an optical readout of mechanical motion by injecting a laser into the WGM through a tapered optical fiber and observing the modulation of the light intensity (or phase) due to the change in the mean optical path due to the mechanical displacement.

The mechanical interaction between the two microbottles originates from the finite mode overlap between the RBMs in each microbottle. The strength of the mechanical interaction

Motoki Asano, Hiroshi Yamaguchi, and Hajime Okamoto are with the NTT Basic Research Laboratories, 3-1 Morinosato Wakamiya, Atsugi-shi, Kanagawa 243-0198, Japan. (e-mail: motoki.asano@ntt.com, hrsh.yamaguchi@ntt.com, hajime.okamoto@ntt.com).

can be adjusted by designing the shallow neck in the middle. The shallow neck brings the mechanical modes into the strong coupling regime where the coupling rate is larger than the mechanical damping rate. The strong coupling provides two hybridized mechanical modes (i.e., in-phase and anti-phase modes) which allows coherent signal transduction from the bottom RBM to the top RBM.

The fluid-structure interaction in the bottom microbottle is crucial for detecting the liquid level in this architecture. Note that here we assume that the liquid is an incompressible viscous medium where the mechanical frequency and linewidth are modulated with respect to the viscosity and density of the liquid [16]. The changes in mechanical frequency and linewidth depend on the spatial overlap between the bottom RBM and liquid. Thus, the liquid level is detectable by monitoring the mechanical frequency shift (or linewidth broadening).

Thus, once the water level is changed, the frequency shift in the hybridized mechanical mode is sensitively read out via the optomechanical coupling at the top microbottle.

## III. DEVICE GEOMETRY AND SETUP

### A. Device fabrication

A silica glass processor (Thorlabs GPX3400) was used to fabricate the TMBR on a silica fiber (Thorlabs SM1500G80) whose diameter was 80 μm. Three shallow necks were formed via the heat-and-pull technique [18],[19]. The overlap-tapering method, where the tapering areas have a finite overlap along the fiber-axis direction, was used to downsize the TMBR [17]. The maximum diameter was 68 μm, the neck diameter was 62 μm, and the length between the two necks was 400 μm (i.e., the total device length was about 800 μm). Figure 1(c) shows an optical microscope image of the fabricated TMBR.

### B. Setup and optomechanical properties

An external cavity diode laser (ECDL; TOPTICA Photonics CTL1550) was used to evaluate the optomechanical properties of the TMBR. This was done by adjusting its power and polarization with a variable optical attenuator (VOA) and polarization controller (PC) [Fig. 2(a)]. The laser light passed into the WGM of the TMBR via a tapered optical fiber whose diameter was about 1.5 μm. The transmitted light was split into two: 10% was sent to the avalanche photodiode (APD) to monitor the optical transmission spectra; and 90% was sent to the balanced detector (BD) after mixing the polarizations with a PC and a polarizing beam splitter (PBS). To perform balanced homodyne detection, the PBS mixed two polarization components: the off-resonant component in the WGM that was used as a local oscillator and the on-resonant component which includes the optomechanical signal passing through the TMBR [20].

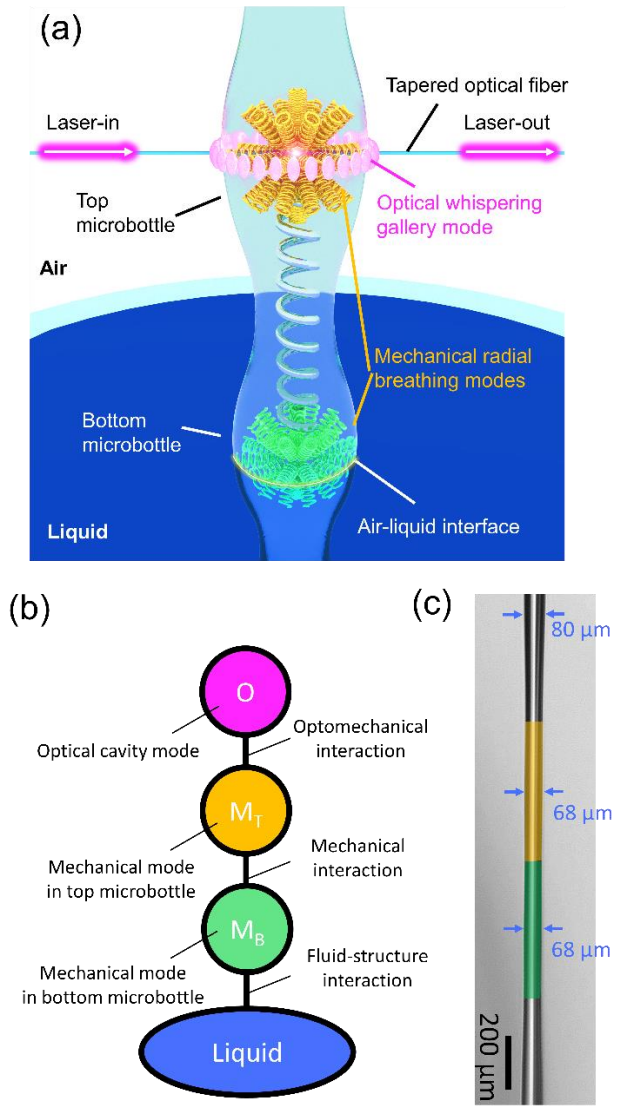

Fig. 1 (a) Illustration of the TMBR for use as a liquid level meter. (b) Schematic diagram of the opto-mechano-fluid interactions in the TMBR. (c) Optical microscope image of the TMBR. The yellow and green shaded area corresponds to a microbottle structure.

The optical properties of the TMBR were evaluated via the optical transmissions monitored by the oscilloscope. The transmission spectrum of the WGM in the TMBR was gathered by scanning the laser frequency [Fig. 2(b)]. A linewidth of 60 MHz corresponding to an optical Q factor of $3.2 \times 10^6$ was achieved. This high optical Q factor is comparable to those of standard optical microbottle resonators [21]-[24].

The mechanical properties of the TMBR were evaluated by observing mechanical motion in the thermal fluctuations via optomechanical coupling. Here, the laser frequency was adjusted to the slope of the resonance where the optomechanical transduction coefficient takes the maximum value, and the laser power was set to about 100 μW. The observed power spectral density (PSD) of the thermal fluctuation is shown in Fig. 2(c). The two peaks correspond to the hybridized mechanical modes where the lower (higher) frequency mode corresponds to the in-phase (anti-phase) hybridized mode. The resonance frequency was around 56.7 MHz and the mechanical Q factor was about 3500. The two peaks were well fitted to an analytical function derived from coupled mode theory (the black dotted line). The coupling strength between the two mechanical modes was estimated to be $g_M/2\pi = 225$ kHz, which is much larger than the linewidth of the mechanical modes of 16.2 kHz. Thus, the TMBR operated in the strong coupling regime for the two mechanical modes.

The optomechanical interaction can be quantified by the optomechanical coupling rate, $g_{OM} = x_{zpf}\partial\omega_{cav}/\partial x$, where $x_{zpf}$ is the zero-point fluctuation of the mechanical mode, $\omega_{cav}$ is the optical cavity resonance frequency, and $x$ is the mechanical displacement. The optomechanical coupling rate, $g_{OM}/2\pi = 0.4$ kHz, was estimated by measuring the thermal fluctuation and comparing the PSD with that of a calibration tone with a known phase modulation from a phase modulator (PM) [25].

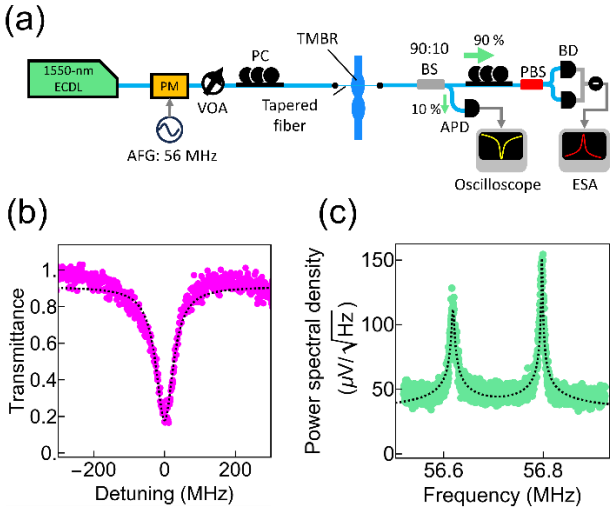

positioner [Fig. 3(a)]. Figure 3(b) shows the PSD of the thermal fluctuation in the coupled mechanical modes with different immersion depths. Mechanical frequency shifts can be seen as well as the linewidth broadening due to the viscous damping [10],[16]. Moreover, the lower frequency mode shows a dramatic change in its mechanical properties while the higher frequency mode shows a small change. This implies that the lower frequency mode has the dominant spatial distribution in the bottom microbottle due to the initial frequency difference between the two mechanical modes. Because the lower frequency mode is much more sensitive than the higher frequency mode, we will focus on detection of the water level by using the lower frequency mode in what follows.

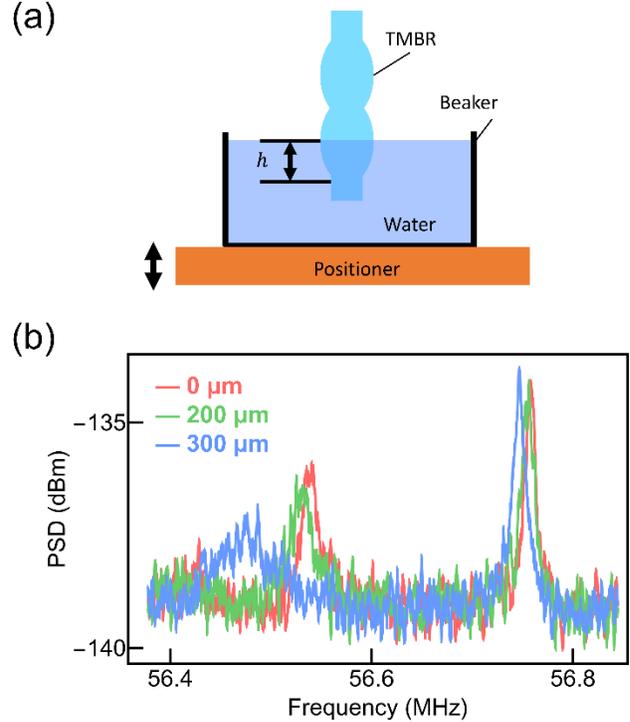

Fig. 2 (a) Schematic diagram of the experimental setup for evaluating optomechanical properties of the TMBR. ECDL: external cavity diode laser, AFG: arbitrary function generator, PM: phase modulator, VOA: variable optical attenuator, PC: polarization controller, BS: beam splitter, APD: avalanche photodiode, PBS: polarizing beam splitter, BD: balanced detector, ESA: electric spectrum analyzer. (b) Optical transmission spectrum of the TMBR. The dotted line is the fitting with a Lorentz function. (c) Power spectral density of the thermal fluctuation of the hybridized mechanical modes in the TMBR. The dotted line is the theoretical fitting following coupled mode theory.

## IV. WATER LEVEL DETECTION

### A. Water level detection via spectral measurement

The water level with respect to the TMBR can be detected by monitoring the mechanical frequency shift with respect to the immersion depth, $h$. The TMBR was gradually immersed in the water in a beaker by controlling the immersion depth with a

Fig. 3 (a) Schematic diagram of experimental setup for water-level detection. $h$ denotes the immersion depth. (b) Power spectral density of thermal fluctuations at different immersion depths.

The resolution for the water level can be quantified by

$$\delta h = \left(\frac{\partial f}{\partial h}\right)^{-1} \delta f_M \qquad (1)$$

where $\partial f/\partial h$ is the water-level responsivity with respect to the mechanical frequency, and $\delta f_M$ is the mechanical frequency resolution which depends on the frequency tracking scheme, as discussed in regard to various mechanical systems [26],[29]. To estimate the resolution, we experimentally determined the water-level responsivity with respect to the immersion depth. The responsively was measured by monitoring the spectra in the lower frequency mode while changing the immersion depth and the mechanical frequency shift [Fig. 4(a) and (b)]. The responsivity increased with increasing immersion depth because the mechanical frequency shift $\Delta f_M$ monotonically and nonlinearly decreased with immersion depth. This result shows that the deeper the immersion depth is, the better the resolution becomes.

On the other hand, the mechanical linewidth increased with increasing immersion depth [Fig. 4(c)]. As the linewidth increased, the signal-to-noise ratio of the mechanical motion decreased, resulting in decreased resolution. In the following experiment, we set the immersion depth to be around 250 μm in consideration of this resolution trade off.

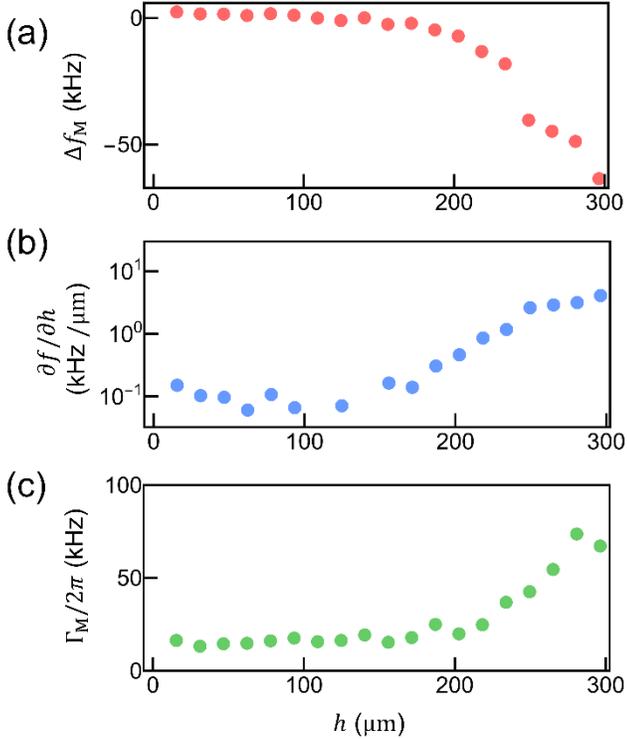

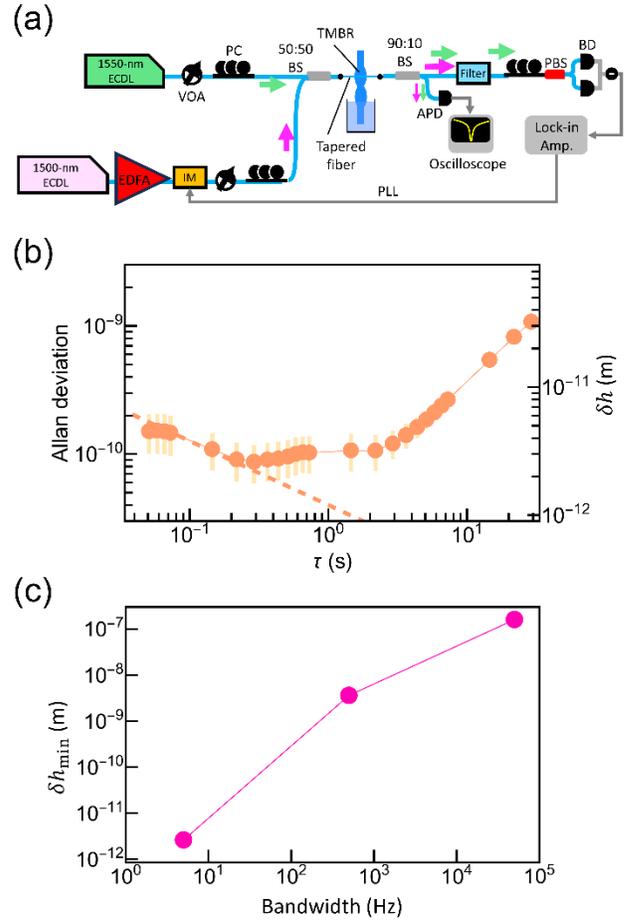

Fig. 4 (a) Mechanical frequency shift, $\Delta f_M$, (b) water-level responsivity, $\partial f / \partial h$, and (c) mechanical damping rate, $\Gamma_M$ with respect to the immersion depth of $h$.

### B. Estimation of resolution under phase-locked loop

The phase-locked loop (PLL) is a powerful way to perform real-time tracking of mechanical frequency. The resolution for the water level can be written as

$$\delta h = \left(\frac{\partial f}{\partial h}\right)^{-1} f_M \sigma_A \equiv c_M \sigma_A \quad (2)$$

where $\sigma_A$ is the Allan deviation given by

$$\sigma_A = \frac{1}{f_0} \sqrt{\frac{\langle (f_M(t+\tau) - f_M(t))^2 \rangle}{2}} \quad (3)$$

with an integration time of $\tau$ and average mechanical frequency of $f_0$ [30]. To implement the PLL, the second ECDL (TOPTICA Photonics, CTL1500) was used to drive the mechanical motion by setting its optical power to several mW with an erbium-doped fiber amplifier (EDFA) and modulating its intensity by the mechanical resonance frequency with an intensity modulator (IM: Thorlabs LNA2322) [Fig. 5(a)]. The probe light from the first ECDL was filtered out with an optical filter after the tapered optical fiber for probing the mechanical motion. The detected signal was sent to a lock-in amplifier (Zurich Instruments UHFLI) and fed back to the IM. The factor around $h \sim 250$ μm was estimated to be $c_M = 3.0 \times 10^{-2}$ m.

Fig. 5 (a) Schematic diagram of the experimental setup for the PLL measurement. EDFA: erbium-doped fiber amplifier, IM: intensity modulator. (b) Allan deviation with respect to the integration time $\tau$ with a 5-Hz bandwidth. The dotted line corresponds to the $\tau^{-1/2}$-dependence. The error bars show the standard deviation of Allan deviation in each $\tau$. (c) Best resolution for the water level with respect to different PLL bandwidths.

By tracking the mechanical frequency via the PLL, the Allan deviation was evaluated with respect to the integration time $\tau$ [Fig. 5(b)]. Here, the bandwidth of the low pass filter in the lock-in amplifier was set to 5 Hz, which was the minimum achievable bandwidth in our setup. Apparently, increasing the integration time decreased the Allan deviation to around 0.05 to 0.2 seconds because the white noise contribution was dominant in this regime [$\tau^{-1/2}$-dependence of the white noise is shown by the dotted line in Fig. 5(b)]. On the other hand, longer integration times of more than 3 seconds reflected worse deviations due to frequency drift in the measurement setup. The corresponding resolution for the water level was estimated by multiplying $c_M$ and the Allan deviation [see the right vertical axis in Fig. 5(b)]. As a result, the best resolution for the water level was found to be $\delta h_{\min} = (2.6 \pm 0.9) \times 10^{-12}$ m for the 5-Hz bandwidth.

To capture water level fluctuations faster than 5 Hz, the bandwidth needs to be increased appropriately. We examined the highest resolution for the water level under the PLL with three different bandwidths, 5, 500, and 50000 Hz [Fig. 5(c)].

Apparently, a bandwidth-resolution trade off exists, wherein increasing the bandwidth reduces the resolution due to the short integration inside the lock-in amplifier.

## V. Discussion

The estimated resolution for the water level, especially in the PLL measurement, is much better than the previously reported liquid level meter based on fiber-technologies whose resolution is the order of the optical wavelengths [31]-[37]. This high resolution originates from the combination of the small mode volume of the mechanical modes and the ultrasensitive displacement measurement of the cavity optomechanical setup. This highly sensitive liquid level meter has the potential to explore the tiny fluctuations of air-liquid and liquid-liquid interfaces. In particular, the optomechanical setup is independent of the optical properties of the liquid, so this technique can be applied to various chemical and biological species.

The dynamic range of the liquid-level meter is restricted by the length of the bottom microbottle resonator (400 μm in the current TMBR). By using the multiply chained microbottle resonator (CMBR) recently reported by the authors [38], the length of the sensor part of the microbottle resonator can be greatly extended to a few centimeters. Although a trade-off exists between the dynamic range and the resolution, the CMBR architecture allows us to appropriately adjust the dynamic range and resolution by designing the number of the chained microbottles.

The estimation of the resolution for the water level assumes a linear mechanical frequency response to the liquid level. However, there might be further contributions that result in strong nonlinearities, such as surface tension at the air-liquid interface and wettability of the glass surface depending on the type of liquid. Further investigations, including one on the detailed fluid-structure interaction, will be needed to improve the accuracy of the estimation for the resolution.

## VI. Conclusion

We demonstrated liquid-level measurements with a twin-microbottle resonator, which enables the air-liquid interface to be used to detect the liquid level. A water-level responsivity was estimated by monitoring the thermal fluctuations of the coupled mechanical mode with immersing the device into water. Furthermore, a phase-locked loop measurement provided a resolution of $2.6 \pm 0.9$ pm at the minimum bandwidth of 5 Hz. This unique cavity optomechanical architecture that takes advantage of the liquid's shape will lead to novel sensor applications for various chemical and biological targets and soft matter.

## Acknowledgments

This work was supported by JSPS KAKENHI (21H01023, 23H05463).